\documentclass[fleqn,usenatbib,useAMS,letters]{mnras}

\newcommand{\hei}{He~I} 

\newcommand{\lyalph}{Lyman $\alpha$}

\newcommand{\gaia}{\emph{Gaia}}
\newcommand{\ktwo}{\emph{K2}}

\newcommand{\hst}{\emph{HST}}

\newcommand{\msun}{M$_{\odot}~$}

\newcommand{\mearth}{$M_\oplus$~}

\newcommand{\rearth}{$R_{\oplus}$}

\usepackage{newtxtext,newtxmath,lineno}

\usepackage[T1]{fontenc}
\usepackage{ae,aecompl}

\usepackage{graphicx}	
\usepackage{amsmath}	
\usepackage{amssymb}	

\title[]{Zodiacal Exoplanets in Time. XI. The Orbit and Radiation Environment of the Young M Dwarf-Hosted Planet K2-25b}

\author[Gaidos et al.]{
E. Gaidos\thanks{E-mail: gaidos@hawaii.edu}$^{1}$, T. Hirano$^{2}$,
D. J. Wilson$^{3}$, K. France$^{4}$, K. Rockcliffe$^{5}$, E. Newton$^{5}$, \newauthor 
G. Feiden$^{6}$, 
V. Krishnamurthy$^{2}$,
H. Harakawa$^{7,8}$, 
K. W. Hodapp$^{9}$, 
M. Ishizuka$^{10}$, 
\newauthor
S. Jacobson$^{9}$,
M. Konishi$^{11}$, 
T. Kotani$^{8,12,13}$,
T. Kudo$^{7,8}$, 
T. Kurokawa$^{8,14}$,
\newauthor
M. Kuzuhara$^{8,12}$, 
J. Nishikawa$^{8,12,13}$,
M. Omiya$^{8,12}$,
T. Serizawa$^{14}$,
M. Tamura$^{8,10,12}$,
\newauthor
A. Ueda$^{12}$,
S. Vievard$^{7,8}$
\\
$^{1}$Department of Earth Sciences, University of Hawai'i at M\={a}noa, Honolulu, HI  96822, USA\\
$^{2}$Department of Earth and Planetary Sciences, Tokyo Institute of Technology, 2-12-1 Ookayama, Meguro-ku, Tokyo 152-8551, Japan\\
$^{3}$McDonald Observatory, University of Texas at Austin, Austin, TX 78712 USA\\
$^{4}$Laboratory for Atmospheric and Space Physics, University of Colorado, Boulder, CO 80309, USA\\
$^{5}$Department of Physics and Astronomy, Dartmouth College, Hanover, NH 03755, USA\\
$^{6}$Department of Physics and Astronomy, University of North Georgia, Dahlonega, GA 30597 USA\\
$^{7}$Subaru Telescope, 650 N. Aohoku Place, Hilo, HI 96720, USA\\
$^{8}$Astrobiology Center, NINS, 2-21-1 Osawa, Mitaka, Tokyo 181-8588, Japan\\
$^{9}$University of Hawaii, Institute for Astronomy, 640 N. Aohoku Place, Hilo, HI 96720, USA\\
$^{10}$Department of Astronomy, Graduate School of Science, The University of Tokyo, 7-3-1 Hongo, Bunkyo-ku, Tokyo 113-0033, Japan\\
$^{11}$Faculty of Science and Technology, Oita University, 700 Dannoharu, Oita 870-1192, Japan\\
$^{12}$National Astronomical Observatory of Japan, NINS, 2-21-1 Osawa, Mitaka, Tokyo 181-8588, Japan\\
$^{13}$Department of Astronomy, School of Science, The Graduate University for Advanced Studies (SOKENDAI), 2-21-1 Osawa, Mitaka, Tokyo, Japan\\
$^{14}$Institute of Engineering, Tokyo University of Agriculture and Technology, 2-24-16, Nakacho, Koganei, Tokyo, 184-8588, Japan
}

\date{Accepted XXX. Received YYY; in original form ZZZ}

\pubyear{2020}

\hypersetup{draft}
\begin{document}
\label{firstpage}
\pagerange{\pageref{firstpage}--\pageref{lastpage}}
\maketitle

\begin{abstract}
M dwarf stars are high-priority targets for searches for Earth-size and potentially Earth-like planets, but their planetary systems may form and evolve in very different circumstellar environments than those of solar-type stars.  To explore the evolution of these systems, we obtained transit spectroscopy and photometry of the Neptune-size planet orbiting the $\approx$650 Myr-old Hyades M dwarf K2-25.  An analysis of the variation in spectral line shape induced by the Doppler ``shadow" of the planet indicate that the planet's orbit is closely aligned with the stellar equator ($\lambda=-1.7_{-3.7}^{+5.8}$ deg), and that an eccentric orbit found by previous work could arise from perturbations by another planet on a co-planar orbit.  We detect no significant variation in the depth of the He I line at 1083\,nm during transit.  A model of atmospheric escape as a isothermal Parker wind with a solar composition show that this non-detection is not constraining compared to escape rate predictions of $\sim$0.1 \mearth\,Gyr$^{-1}$; at such rates, at least several Gyr are required for a Neptune-like planet to evolve into a rocky super-Earth.    
\end{abstract}




\section{Introduction}
\label{sec:intro}

M dwarfs are numerous hydrogen-burning stars with comparatively low masses, small radii, and low luminosities that promote the detection of Earth-size and potentially Earth-like planets in compact circumstellar habitable zones.  However, the formation and environments of planets on short-period orbits around these stars could differ substantially from those of true Earth analogs.  M dwarfs more frequently host compact multi-planet systems \citep{Muirhead2015} and host more close-in planets \citep{Mulders2015}, possibly a consequence of differences in the structure and evolution of their circumstellar disks \citep{Kastner2016,Gaidos2017b}.  M dwarfs have elevated X-ray and UV emission (XUV) relative to their bolometric (total) luminosities, and exhibit a different evolution in the rotation and magnetic activity responsible for this emission \citep{Matt2015}.  M dwarf stars have prolonged pre-main sequence phases compared to their solar-mass counterparts.  Such differences could drive planets and their atmospheres on divergent evolutionary pathways as a result of the runaway greenhouse effect \citep{Luger2015}, photo-dissociation of molecules \citep{Tian2014}, and escape of primordial H/He atmospheres to space \citep{Owen2019}.

The evolution of planets can be investigated by observations of systems in nearby young stellar clusters and co-moving groups.  For transiting planets it is possible to make unambiguous radius and mass estimates to determine bulk density and constrain composition, to measure the atmosphere via transmission spectroscopy, and to constrain the orbit via measurements of the duration and times of transit as well as the spin-orbit obliquity via the Rossiter-McLaughlin effect \citep{Triaud2017}.

2MASS J04130560+1514520 is a member of the $\approx$650 Myr-old Hyades cluster which hosts a Neptune-size planet on a 3.45 day transiting orbit.  The planet was discovered in \ktwo{} data \citep{Mann2016a} and designated K2-25b.  Along with the K2-126 system \citep{Mann2018}, it stands because of its young age and proximity ($\approx$50 pc). \citet{Thao2020} compared the transit depth as measured with \emph{Spitzer} 4.5 $\mu$m observations to that measured with \ktwo{} data ($\sim0.8\mu$m) to rule out a solar-composition, H-rich atmosphere.  Precise measurement of the transit duration, combined with a \gaia{}-improved estimates of stellar properties estimation, indicates an orbital eccentricity $e>0.2$,  suggestive of perturbation by a companion planet or star.  An interaction like the Kozai-Lidov resonance would also have left the planet on a highly-inclined orbit.  The host star is rapidly rotating (1.88-day period) and has significant rotational variability indicative of star spots, and thus magnetic activity.  K2-25b's proximity to the star (0.035\,au) thus makes its atmosphere vulnerable to XUV-driven escape.     

To further constrain the orbit and search for atmospheric escape from K2-25b, we made spectroscopic and photometric observations of a transit on UT 13 October 2019 to obtain information on the change in line shape due to the Rossiter-McLaughlin effect ("Doppler tomography") and absorption in the 1083\,nm line of metastable ``triplet" (ortho) \hei{} \citep{Oklopvcic2018}.  We also use \emph{HST} observations of an earlier transit (Rockcliffe et al., in prep.) to constrain the UV emission from the star.  
\vspace{-0.25in}

\section{Observations and Data Reduction}
\label{sec:observations}

\emph{Infrared spectroscopy:}  Seventy-seven spectra of K2-25 were obtained with the IRD infrared echelle spectrograph \citep{2012SPIE.8446E..1TT, Kotani2018} on the Subaru telescope on Maunakea over a 430-day interval beginning 17 August 2018.  IRD covers 970-1730\,nm with $\lambda/\Delta \lambda \approx 70,000$. Twenty-nine spectra were obtained over a 100-min interval, including the 40 min-long transit of K2-25b, on the night of UT 13 October 2019.  Four and 15 spectra were obtained during the previous and following nights, respectively.  Integration times were $5-$min during and near the transit, and 10\,min otherwise.  Using the \texttt{IRAF} echelle package \citep{Tody1986} and custom software, we extracted one-dimensional spectra after flat-fielding and scattered light subtraction.  Wavelengths were calibrated using the comparison spectra of the Th-Ar lamp taken during the run. Typical signal-to-noise (S/N) ratios at 1000\,nm were $20-25$ and $25-35$ per pixel for 300- and 600-sec integrations, respectively.

\emph{Photometry:} Continuous imaging of K2-25 was obtained for 14.5 hours starting 12 hours before transit with the LCOGT 0.4-meter telescopes and  SBIG CCD detectors at the Tenerife (TFN) and McDonald Observatory (ELP) sites.  Each 300-second integration was made through a Sloan $r$' filter.  Images were automatically processed using the {\tt BANZAI} pipeline \citep{McCully2018}.   To remove atmospheric effects, a reference lightcurve was constructed using a set of 130 stars that were iteratively selected for low variability with respect to an aggregate mean.  TFN data are of low photometric quality due to weather and are not presented here.

\emph{UV spectroscopy:} We observed K2-25 with \hst{} and the Space Telescope Imaging Spectrometer \citep[STIS,][] {Riley2017} during visits on 23 March and 31 October 2017.  During each visit, 8 integrations were obtained with the G140M grating and the far-ultraviolet multi-anode micro-channel array (FUV-MAMA) in TIME-TAG mode.  Four of the 16 integrations were 1993 sec and the remainder were 2054 sec, spanning the observable window of the \hst{} orbit.  These data were reduced using the {\tt CALSTSIS} pipeline. The background, dominated by geocoronal emission at 1215\AA, is negligible for the wavelength range of concern here.   The fluxes in the N\,V doublet (1238.821 and 1242.804\AA) were calculated for each exposure using Eqn. 1 from \citet{France2018}. The second visit suffered a large wavelength offset and required a correction of +0.3\AA.  Examination of the N\,V lightcurves identified one integration potentially contaminated by a flare during each visit and these were excluded from further analysis.  The spectral continuum was estimated by calculating the mean flux in 1\AA{} bins on either side of each line and linearly interpolating.   Fluxes between the two visits did not significantly differ and the total N\,V line flux averaged over the two visits is $2.8 \pm 0.3 \times 10^{-16}$ ergs sec$^{-1}$ cm$^{-2}$. 
\vspace{-0.25in}

\section{Analysis and Results}
\label{sec:analysis}

\begin{figure}
	\includegraphics[width=\columnwidth]{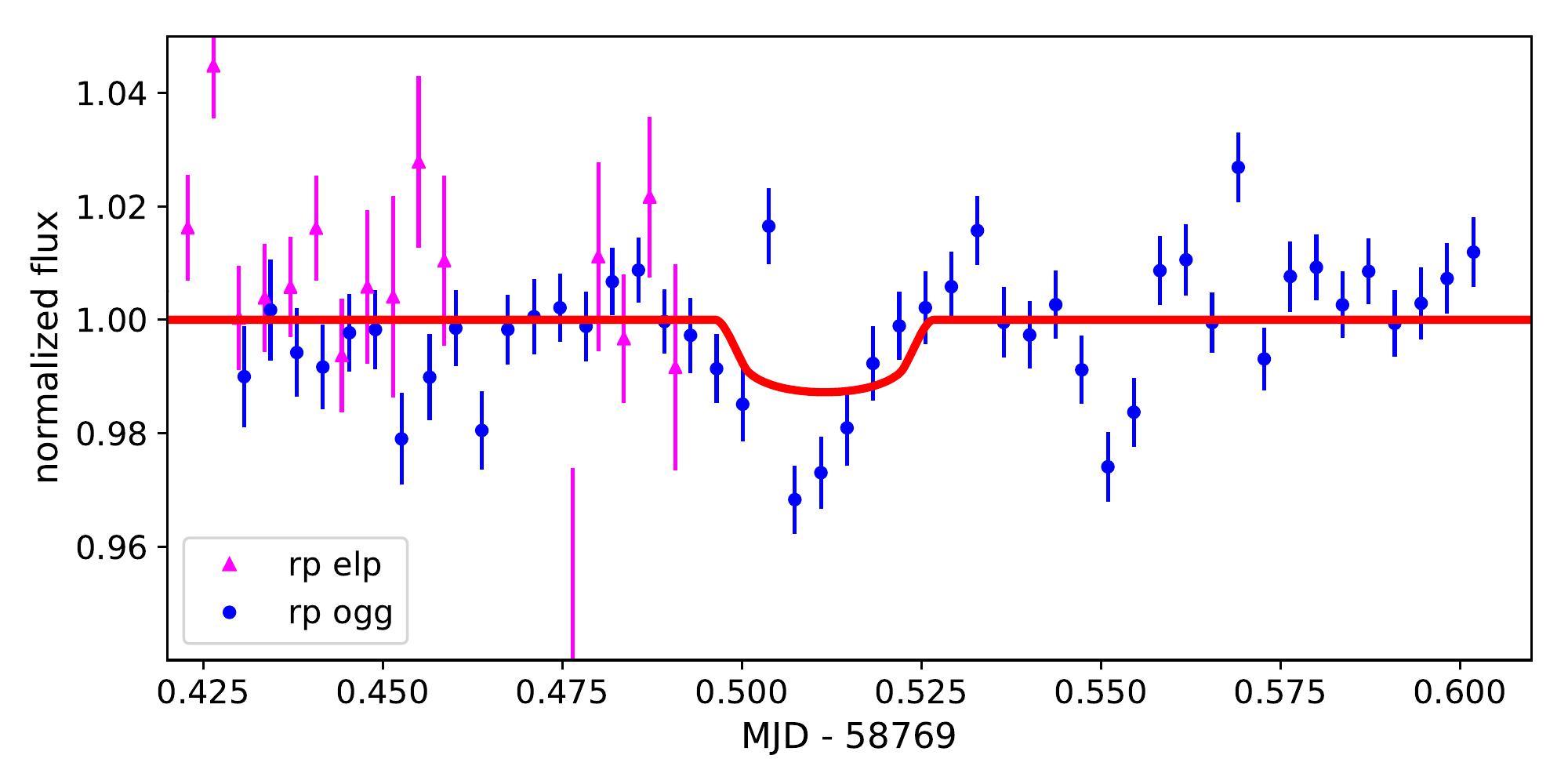}
    \caption{Top: $r$-band lightcurve of K2-25 from LCOGT McDonald (ELP) and Haleakala (OGG) 0.4-m telescopes.  The red curve is a {\tt PyTransit} model \citep{Parvianen2015} using the parameters of \citet{Thao2020}.}  
    \label{fig:loc}
\end{figure}

\subsection{Rossiter-McLaughlin effect and Doppler-shadow analysis}
\label{sec:rm}

Our LCOGT observations confirm that the transit occurred approximately as predicted by the ephemeris of \citet{Thao2020} (Fig. \ref{fig:loc}).  The transit appears 6\, min earlier than predicted ($\pm$1\,min) and significantly deeper, most likely due to systematics in the photometry.  To measure the spin-orbit obliquity, we analyzed the mean line profiles in IRD spectra obtained during the transit.  Following \citet{2020ApJ...890L..27H}, we computed the cross-correlation function (CCF) between each spectrum and a telluric-free template spectrum of the M4 dwarf GJ 699 (Hirano et al. under review). Before computing CCFs, we divided each spectrum by the normalised spectrum of a telluric standard star obtained on the same night. To make a high S/N CCF for K2-25, we combined the normalised CCFs after correcting for the barycentric motion of Earth.  This mean ``out-of-transit" CCF was then subtracted from individual normalised CCFs to visualise the instantaneous variation of the line profile. 

\begin{figure*}
	\includegraphics[width=16cm]{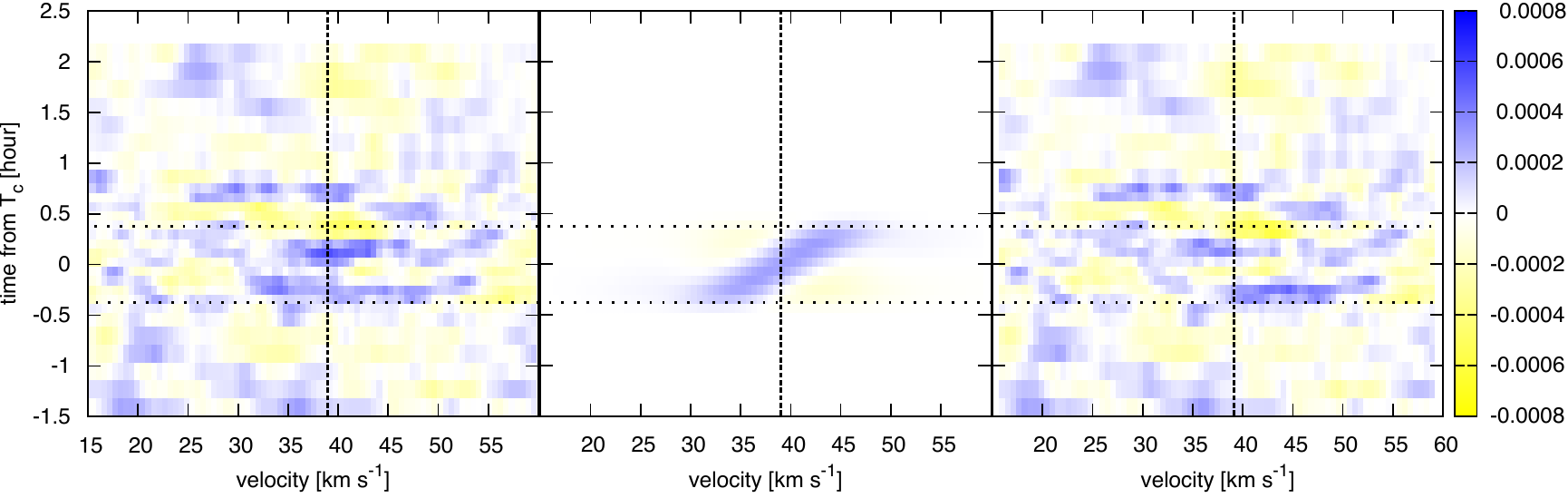}
    \caption{Residual CCF maps based on IRD spectra of K2-25 obtained during a transit of ``b" on UT 13 October 2019. The left and middle panels are the observed residual CCF and best-fit model, respectively. The right panel is the residual map after subtracting the best-fit model. The vertical dashed line marks the centre of the mean CCF ($\approx 39$ km s$^{-1}$) at the systemic radial velocity (RV) of K2-25, and the horizontal dotted lines indicate transit ingress and egress.}  
    \label{fig:DT1}
\end{figure*}

The left panel of Fig. \ref{fig:DT1} plots the residual CCFs vs. time.  The loss of the spectral contribution from the part of the stellar disk occulted by the planet should appear as a moving feature (blue band) during the transit (between the horizontal dotted lines), but is not apparent due to low S/N.  To estimate the projected obliquity $\lambda$ we compared the observed residual CCFs to a model based on mock IRD spectra with different values of the stellar rotation velocity $v\sin i$ (4-11 km s$^{-1}$) and planet position on the stellar disk (see \citet{2020ApJ...890L..27H} for more details). We fit these models to the observed residual CCF map with a Markov Chain Monte Carlo (MCMC) analysis without a prior on $v\sin i$.  However, this MCMC fit did not converge in $v\sin i$, most likely due to low S/N in the CCF around the transit. We thus imposed a prior on $v\sin i$ based on the observed mean CCF.  A comparison between the observed and model CCFs gave $v\sin i=7.71\pm 0.29$ km s$^{-1}$, which agrees with that derived by \citet{Mann2016a}.  Using this value as a Gaussian prior, we refit the observed residual CCFs by MCMC. In the analysis, we also allowed the scaled semi-major axis ($a/R_*$), impact parameter $b$, and mid-transit time $T_c$ to vary with Gaussian priors based on the values in \citet{Thao2020}. 

\begin{figure}
	\includegraphics[width=\columnwidth]{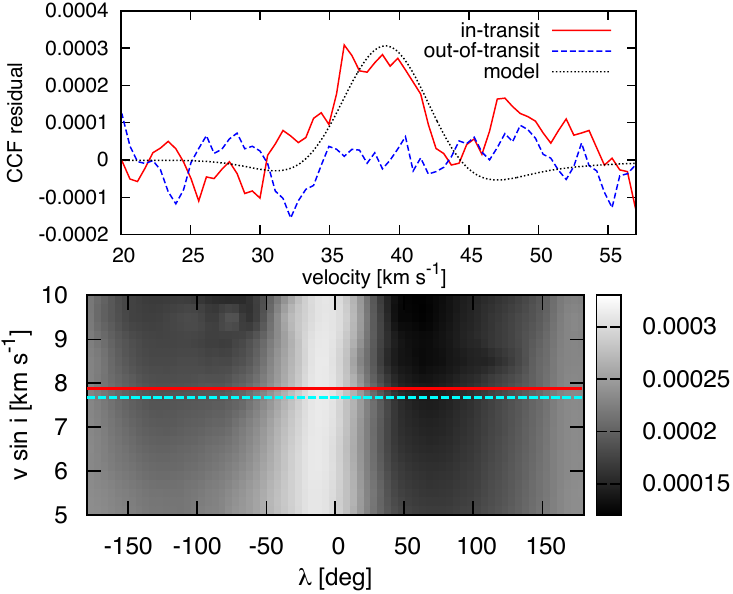}
    \caption{Top: Average residual CCFs during (red) and outside (blue) the transit of K2-25b. The same number of spectra were used in each profile. The black dotted line is the best-fit model. Bottom: Grey scale map of the peak height of the combined in-transit CCF for different sets  of $\lambda$ and $v\sin i$. The solid (red) and dashed (cyan) horizontal lines are the equatorial rotation velocity estimated from $P_\mathrm{rot}$ and $v\sin i$ from our spectroscopy, respectively. }  
    \label{fig:ccf_peak}
\end{figure}

Our MCMC fit yielded an obliquity of $\lambda=-1.7_{-3.7}^{+5.8}$ deg, consistent with spin-orbit alignment.   The middle panel of Fig.\ref{fig:DT1} shows the best-fit theoretical model for the observed CCF map.   To validate this detection, we confirmed that the depth of the feature is consistent with predictions, Doppler-shifting each residual CCF frame during the transit so that the shifted CCF has the expected peak at the systemic RV in Fig.\ref{fig:DT1}, and averaging the shifted CCFs.  The red solid line in the upper panel of Fig.\ref{fig:ccf_peak} plots the combined residual CCF during the transit; the observed peak height is consistent with that of the combined theoretical CCF derived from the same spectra.  For comparison, the blue dashed line in the same panel shows the combined out-of-transit CCFs using the same number of closest frames to the transit. The observed peak height of the combined in-transit CCF is $5-6$ times higher than the scatter of the combined out-of-transit CCF ($\approx 0.000054$). 

As a further test, we checked if the combined in-transit CCF also has the highest peak near $\lambda=0$ degrees. To do so, we combined the in-transit residual CCFs, each Doppler-shifted based on a set of ($\lambda$, $v\sin i$) values, and fitted the resulting mean in-transit CCFs by a Gaussian. For each set of ($\lambda$, $v\sin i$) values we recorded the Gaussian height and generated its contour map. The lower panel of Fig.\ref{fig:ccf_peak} show that the highest CCF peak ($>0.0003$) fall in the range $-21 \lesssim \lambda \lesssim +1$ deg, in agreement with $\lambda=-1.7^{+5.8}_{-3.7}$ deg from the direct MCMC fitting above, while the CCF height has a very weak dependence on $v\sin i$ within the simulated range.

\subsection{1083 nm \hei{} line}
\label{sec:hei}

A prominent telluric OH line interfered with the stronger unresolved doublet (1083.025 and 1083.034 nm) in the stellar line of He I in IRD spectra obtained in August 2018 and October 2019, including during the transit.  The same line interfered with the weaker singlet (1082.909 nm) line in spectra obtained during January-March 2019, and for consistency we excluded those data from further analysis.  We estimated non-transit related variability due to rotation and evolution of active regions on the star using the weaker line.  Because the SNR of individual measurements is limited, we determined a best-fit Voigt profile based on a sum of all spectra shifted into the rest-frame of the star.  This profile was then fit to individual spectra to estimate the equivalent width (EW).  We equated the measurement uncertainty with the intra-night variability (20 m\AA.)  Assuming Gaussian-distributed errors, the mean in the 2018 and 2019 intervals are $63 \pm 4$ and $100 \pm 4$ m\AA{}; the difference indicates long-term variation in the stellar line.   

\begin{figure}
	\includegraphics[width=\columnwidth]{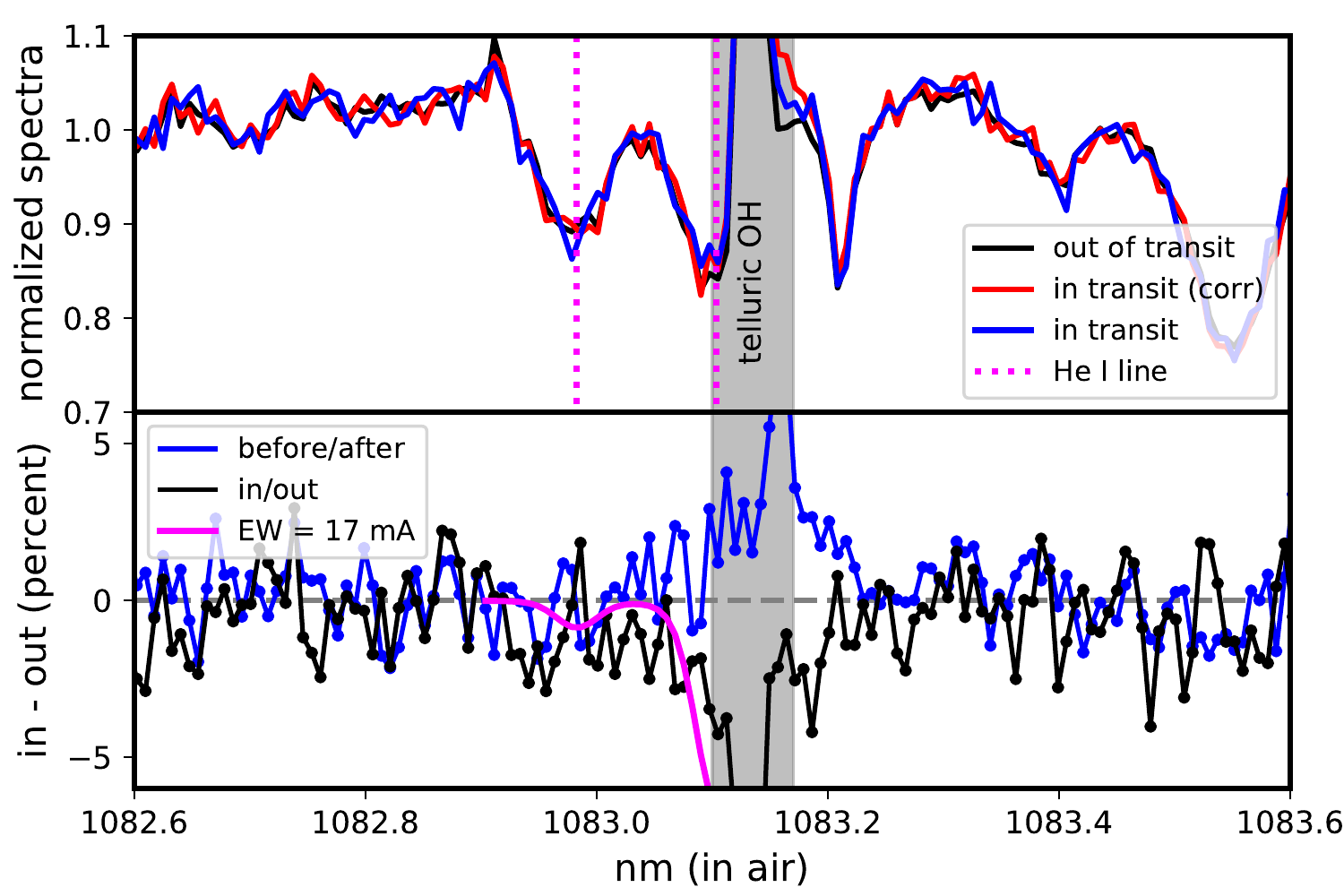}
    \caption{Top: Spectra of K2-25 in the vicinity of the He I line inside and outside of the transit of ``b".  Wavelengths are in air, in the rest frame of the star.  The grey zone contains a strong telluric OH line.  Bottom: The difference spectrum, compared to a model spectrum of a line with a partial (blueward) EW of 17 m\AA, the 99\% confidence detection limit.}  
    \label{fig:hei_line}
\end{figure}

To search for an escaping atmosphere, IRD spectra of K2-25 obtained during the transit (8 spectra) and outside transit but on the same night (21 spectra) were co-added.  The two combined spectra are shown in the top panel of Fig. \ref{fig:hei_line}, and the difference spectrum (in/out minus one) is plotted in the bottom panel, along with the difference between spectra obtained before and after the transit as a comparison.  No transit-associated absorption is apparent.  We calculated a $\chi^2$ between the in/out difference spectrum and a Voigt profile of the line assuming a gas temperature of 10000K (see below), and established a 99\% confidence limit of 17\,m\AA{} on the transit-associated EW over a wavelength range covering the weak singlet line and the uncontaminated blue half of the doublet; this limit is represented by the magenta line in the bottom panel of Fig. \ref{fig:hei_line}.  We also examined the Paschen $\beta$ line at 1.282 $\mu$m, an important indicator of accretion around young stellar objects \citep[e.g.,][]{Yasui2019} in the same fashion, but we saw no significant difference between the spectra inside and outside of transit.

We converted this EW limit into a limit on atmospheric escape using a model of a spherical, isothermal Parker wind with a solar-like composition (H/He = 10.5) and photochemistry as described in \citet{Oklopvcic2018} and \citet{Gaidos2020}.  Densities in such winds are low and the metastable triplet (2$^3$S) state of the \hei\ transition is primarily populated by recombination of He ionized by extreme ultraviolet (EUV) photons with energies $>26.4$ eV ($\lambda < 504$\AA) and primarily depopulated by ionizing near ultraviolet (NUV) photons with energies $>4.8$ eV \AA{} \citep[$\lambda < 2583$,][]{Oklopvcic2018}.  For this reason, interpretation of our observations requires knowledge of the stellar UV irradiation of the planet.  Prior to our \hst\ observations, no UV spectrum of K2-25 was available, nor was K2-25 detected in either X-rays or UV.  To best reproduce both the overall spectral energy distribution of K2-25 and detailed spectral features important to photoionization calculations, we adopted a composite spectrum of the weakly active M3 dwarf GJ 674 produced for the Mega-MUSCLES survey \citep{Froning2019} and scaled the flux densities in different wavelength ranges to match expected values.  The full SED of GJ 674 was created by combining: (a) spectra spanning 1070~--~5700~\AA\ obtained with 9 orbits of observations with \emph{HST} and the Cosmic Origins Spectrograph with the G130M and G230L gratings and STIS with the G140M, G140L, G230L, and G430L gratings; (b) X-ray spectra obtained with a $\approx$15\,ksec \emph{XMM-Newton} integration; and (c) BT-Settl PHOENIX stellar photosphere models \citep{Allard2012}.  EUV flux densities were estimated using the solar active region scaling relations of \citet{Linsky2014}.   The \lyalph\ line profile was reconstructed from the STIS G140M spectra via the techniques detailed in \cite{Youngblood2016}.   

We adjusted the GJ 674 spectrum using estimates of the high-energy emission from K2-25 based partly on its rotation and partly on our \hst{} observations of FUV line emission (Sec. \ref{sec:observations}).  We used an activity-rotation relation formulated in terms of the Rossby number $Ro = P_{\rm ROT}/\tau$, where $P_{\rm ROT}$ is the rotation period (1.88 days) and the convective turnover time $\tau$ for the mass \citep[0.26\msun,][]{Thao2020}, metallicity ([Fe/H] = +0.15), and age \citep[650 Myr][]{Martin2018} of K2-25 was estimated using the standard and magnetic Dartmouth stellar evolution models to be 54 and 59 days, respectively \citep{Feiden2016}.  This yields $Ro = 0.032-0.035$, placing it on the ``saturated" part of the rotation-activity relation with an expected $\log L_x/L_{bol} = -3.05 \pm 0.18$ \citep{Wright2018}.  The corresponding luminosities for ``saturated" stars in the \emph{GALEX} far ultraviolet (FUV, 1350-1750\AA) and near ultraviolet (NUV, 1750-2800\AA) channels are about $3 \times 10^{28}$ ergs sec$^{-1}$ and $9 \times 10^{28}$ ergs sec$^{-1}$, respectively \citep{Ansdell2015}.  We estimated the emission in the \lyalph\ line using the relation with X-rays of \citet{Linsky2013}.  We estimated the EUV emission in the 90-360\AA{} range using the flux in the N\,V line estimated in Sec. \ref{sec:observations} and the FUV-EUV relations described by \citet{France2018}.  Lastly, we estimated fluxes at 360-912 \AA{} in 40 or 100\AA\ intervals using the \lyalph\ flux estimated above and the relations of \citet{Linsky2014}.  Table \ref{tab:irradiance} lists the estimated irradiance of K2-25b in the different wavelength regimes, using the stellar parameters of \citet{Thao2020}.  

{\small
\begin{table}
\begin{center}
\caption{Irradiation of K2-25b \label{tab:irradiance}} 
\begin{tabular}{llll}
\hline\hline
\multicolumn{1}{c}{regime} & \multicolumn{1}{c}{$\lambda\lambda$} & \multicolumn{1}{c}{flux at K2-25b} & \multicolumn{1}{c}{source}\\
\hline
 & \multicolumn{1}{c}{\AA} & \multicolumn{1}{l}{$10^3$\,ergs~s$^{-1}$~cm$^{-2}$} & TW = This Work\\
\hline
{\footnotesize bolom.} & --- & $8900 \pm 300$ & \citet{Thao2020}\\
NUV & 1750-2800 & $4.4 \pm 1.5$ & {\footnotesize TW+\citet{Ansdell2015}}\\
FUV & 1350-1750 & $1.3 \pm 0.6$ & {\footnotesize TW+\citet{Ansdell2015}}\\
Ly~$\alpha$ & 1180-1250 & $5.4 \pm 7.0$ &  {\footnotesize TW+\citet{Linsky2013}}\\
EUV & 90-360 & $1.6 \pm 0.8$ & {\footnotesize TW+\citet{France2018}}\\
EUV & 360-912 & $2.2 \pm 2.1$ & {\footnotesize \citet{Linsky2014}}\\
X-ray & 5-120 & $7.9 \pm 3.4$ & {\footnotesize TW+\citet{Wright2018}}\\
\hline
\end{tabular}
\end{center}
\end{table}
}

\begin{figure}
	\includegraphics[width=\columnwidth]{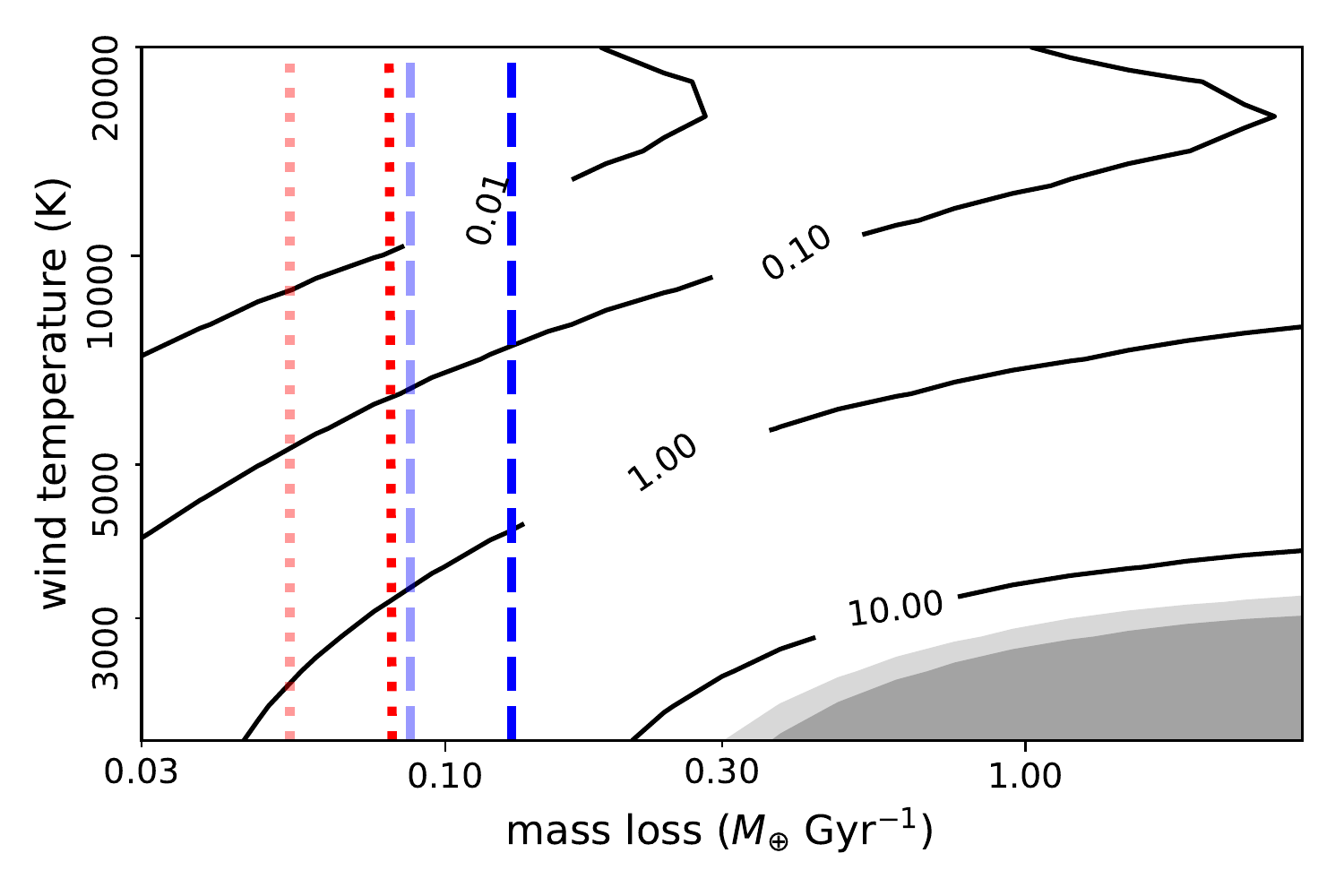}
    \caption{Contours of predicted EW (in m\AA) of the 1083 nm triplet \hei{} line vs. escape rate in \mearth\ Gyr$^{-1}$ ($\approx 2 \times 10^{12}$\,g~sec$^{-1}$) and wind temperature.  Only the weak single line and blue half of the doublet are included.  The grey zones are excluded at 90\% and 99\% confidence by upper limits of 14.5 and 17 m\AA.  The red dotted and blue dashed lines are predictions using  Eqn. \ref{eqn:energy-limited} and the relations of \citet{Kubyshkina2018}, respectively, with the heavy and light version using XUV irradiances with and without the \lyalph\ contribution, respectively.}  
    \label{fig:hei}
\end{figure}

Densities of neutral and ionized H, singlet and triplet neutral He, ionized He, and electrons are tracked.  Line profiles are calculated by integrating from the projected radius of the planet to the Roche radius, accounting for finite optical depth, intrinsic and thermal broadening, and the resolution of the instrument.  The mass of K2-25b is not yet determined so we adopted a planet mass $M_p = 9.7$\, \mearth{} based on its radius of 3.49 \rearth{} \citep{Thao2020} and the mass-radius relation of \citet{Bashi2017}.  Figure \ref{fig:hei} shows the predicted EW of the single line plus the uncontaminated blue half of the doublet as a function of mass loss rate and wind temperature.  

We estimated the energy-limited escape rate by combining the X-ray and photoionizing EUV ($<912$\AA) irradiances in Table \ref{tab:irradiance} as $F_{\rm XUV}$ in \citep{Watson1981,Erkaev2007}:
\begin{equation}
\label{eqn:energy-limited}
\dot{M}_{\rm EL} = \frac{\eta \pi F_{\rm XUV}R_{\rm XUV}^3}{KGM_p},
\end{equation}
where $\eta \sim 10$\% is an efficiency factor \citep{Shematovich2014}, $R_{\rm XUV}$, the effective radius at the level of the atmosphere which the XUV radiation is absorbed and from which escape occurs (here taken to be $\approx R_p$), $G$ is the gravitational constant, $K$ is a correction factor $\approx 1$ that accounts for a finite Roche radius \citep{Erkaev2007}.  We also estimated the escape rate using the empirical relations of \citet{Kubyshkina2018} based on the results of hydrodynamic simulations.  Lyman $\alpha$ photons do not ionize H and He but are resonantly scattered by H\,I and their energy can be absorbed via ionization of, e.g., oxygen \citep{Kockarts2002}.  We therefore made separate estimates with and without the \lyalph\ contribution.  These are all of order 0.1 \mearth\,Gyr$^{-1}$ ( Fig. \ref{fig:hei}.)    Our model shows that, due to the comparatively low EUV emission from small mid-type M type dwarfs such as K2-25, detection of the \hei\ absorption that accompanies expected atmospheric escape rates is challenging \cite[Fig. \ref{fig:hei}, see also ][]{Oklopvcic2019}.  Since elevated EUV drives both H/He escape and triplet \hei\ production, detection of escape is more practical in younger systems where the central star is more rapidly rotating and magnetically active \citep[e.g.,][]{Hirano2020}.

In summary, we find that the orbit of K2-25b is closely aligned with stellar spin, and the consistency between the $v \sin i$ derived here and that from the rotation period plus $R_*$ ($V_\mathrm{eq}=7.90\pm 0.26$ km s$^{-1}$) means that the stellar inclination is near 90 deg.  This geometry suggests that the orbital eccentricity of K2-25b arose from a perturber in the orbital plane, e.g. an undetected planet on an outer orbit, or that the ratio of obliquity damping to orbit circularization timescales $\sim Q'_*/Q'_p \left(M_*/M_p\right)^2$ is $<1$, contrary to theoretical expectation \citep{Matsumura2008,LiG2016}.  Our \hei\ line observations do not usefully constrain atmospheric escape from K2-25b; but at the rates predicted using our estimates of XUV irradiation, the planet could lose $\sim$5\% by mass of H/He over several few Gyr, enough for a transition from a ``Neptune" to a ``super-Earth".

\section*{Acknowledgements}

EG was supported by NASA Grant 80NSSC20K0957 (Exoplanets Research Program) and the German Science Foundation (DFG Research Unit FOR2544.)  IRD was supported by JSPS KAKENHI Grant Numbers 19J11805, 19K14783, 18H05442, 15H02063, and 22000005. We made use of the LCOGT network (DDT award 2019B-003, and data obtained by program \#15071 of the NASA/ESA \emph{HST} from the archive at STScI, operated by AURA, Inc. (NASA contract NAS 5-26555).   We used NASA's Astrophysics Data System Bibliographic Services, the Centre de Donn\'{e}es astronomiques de Strasbourg, NIST's atomic line database, {\tt Astropy} \citep{Astropy2013}, and {\tt Scipy} \citep{Scipy2019}.\\
{\bf Data Availability:}  All data are available from the authors or Subaru SMOKA or STScI HST archives.




\vspace{-0.25in}


\begin{thebibliography}{}
\makeatletter
\relax
\def\mn@urlcharsother{\let\do\@makeother \do\$\do\&\do\#\do\^\do\_\do\%\do\~}
\def\mn@doi{\begingroup\mn@urlcharsother \@ifnextchar [ {\mn@doi@}
  {\mn@doi@[]}}
\def\mn@doi@[#1]#2{\def\@tempa{#1}\ifx\@tempa\@empty \href
  {http://dx.doi.org/#2} {doi:#2}\else \href {http://dx.doi.org/#2} {#1}\fi
  \endgroup}
\def\mn@eprint#1#2{\mn@eprint@#1:#2::\@nil}
\def\mn@eprint@arXiv#1{\href {http://arxiv.org/abs/#1} {{\tt arXiv:#1}}}
\def\mn@eprint@dblp#1{\href {http://dblp.uni-trier.de/rec/bibtex/#1.xml}
  {dblp:#1}}
\def\mn@eprint@#1:#2:#3:#4\@nil{\def\@tempa {#1}\def\@tempb {#2}\def\@tempc
  {#3}\ifx \@tempc \@empty \let \@tempc \@tempb \let \@tempb \@tempa \fi \ifx
  \@tempb \@empty \def\@tempb {arXiv}\fi \@ifundefined
  {mn@eprint@\@tempb}{\@tempb:\@tempc}{\expandafter \expandafter \csname
  mn@eprint@\@tempb\endcsname \expandafter{\@tempc}}}

\bibitem[\protect\citeauthoryear{{Allard}, {Homeier}  \& {Freytag}}{{Allard}
  et~al.}{2012}]{Allard2012}
{Allard} F.,  {Homeier} D.,   {Freytag} B.,  2012, \mn@doi [Philosophical
  Transactions of the Royal Society of London Series A]
  {10.1098/rsta.2011.0269}, \href
  {https://ui.adsabs.harvard.edu/abs/2012RSPTA.370.2765A} {370, 2765}

\bibitem[\protect\citeauthoryear{{Ansdell} et~al.,}{{Ansdell}
  et~al.}{2015}]{Ansdell2015}
{Ansdell} M.,  et~al., 2015, \mn@doi [\apj] {10.1088/0004-637X/798/1/41}, \href
  {http://adsabs.harvard.edu/abs/2015ApJ...798...41A} {798, 41}

\bibitem[\protect\citeauthoryear{{Astropy Collaboration} et~al.,}{{Astropy
  Collaboration} et~al.}{2013}]{Astropy2013}
{Astropy Collaboration} et~al., 2013, \mn@doi [\aap]
  {10.1051/0004-6361/201322068}, \href
  {http://adsabs.harvard.edu/abs/2013A%26A...558A..33A} {558, A33}

\bibitem[\protect\citeauthoryear{{Bashi}, {Helled}, {Zucker}  \&
  {Mordasini}}{{Bashi} et~al.}{2017}]{Bashi2017}
{Bashi} D.,  {Helled} R.,  {Zucker} S.,   {Mordasini} C.,  2017, \mn@doi [\aap]
  {10.1051/0004-6361/201629922}, \href
  {https://ui.adsabs.harvard.edu/abs/2017A&A...604A..83B} {604, A83}

\bibitem[\protect\citeauthoryear{{Erkaev}, {Kulikov}, {Lammer}, {Selsis},
  {Langmayr}, {Jaritz}  \& {Biernat}}{{Erkaev} et~al.}{2007}]{Erkaev2007}
{Erkaev} N.~V.,  {Kulikov} Y.~N.,  {Lammer} H.,  {Selsis} F.,  {Langmayr} D.,
  {Jaritz} G.~F.,   {Biernat} H.~K.,  2007, \mn@doi [\aap]
  {10.1051/0004-6361:20066929}, \href
  {https://ui.adsabs.harvard.edu/abs/2007A&A...472..329E} {472, 329}

\bibitem[\protect\citeauthoryear{{Feiden}}{{Feiden}}{2016}]{Feiden2016}
{Feiden} G.~A.,  2016, \mn@doi [\aap] {10.1051/0004-6361/201527613}, \href
  {https://ui.adsabs.harvard.edu/abs/2016A&A...593A..99F} {593, A99}

\bibitem[\protect\citeauthoryear{{France}, {Arulanantham}, {Fossati}, {Lanza},
  {Loyd}, {Redfield}  \& {Schneider}}{{France} et~al.}{2018}]{France2018}
{France} K.,  {Arulanantham} N.,  {Fossati} L.,  {Lanza} A.~F.,  {Loyd}
  R.~O.~P.,  {Redfield} S.,   {Schneider} P.~C.,  2018, \mn@doi [\apjs]
  {10.3847/1538-4365/aae1a3}, \href
  {https://ui.adsabs.harvard.edu/abs/2018ApJS..239...16F} {239, 16}

\bibitem[\protect\citeauthoryear{{Froning} et~al.,}{{Froning}
  et~al.}{2019}]{Froning2019}
{Froning} C.~S.,  et~al., 2019, \mn@doi [\apjl] {10.3847/2041-8213/aaffcd},
  \href {https://ui.adsabs.harvard.edu/abs/2019ApJ...871L..26F} {871, L26}

\bibitem[\protect\citeauthoryear{{Gaidos}}{{Gaidos}}{2017}]{Gaidos2017b}
{Gaidos} E.,  2017, \mn@doi [\mnras] {10.1093/mnrasl/slx063}, \href
  {http://adsabs.harvard.edu/abs/2017MNRAS.470L...1G} {470, L1}

\bibitem[\protect\citeauthoryear{{Gaidos} et~al.,}{{Gaidos}
  et~al.}{2020}]{Gaidos2020}
{Gaidos} E.,  et~al., 2020, \mn@doi [\mnras] {10.1093/mnras/staa918}, \href
  {https://ui.adsabs.harvard.edu/abs/2020MNRAS.tmp..180G} {}

\bibitem[\protect\citeauthoryear{{Hirano} et~al.,}{{Hirano}
  et~al.}{2020a}]{Hirano2020}
{Hirano} T.,  et~al., 2020a, arXiv e-prints, \href
  {https://ui.adsabs.harvard.edu/abs/2020arXiv200613243H} {p. arXiv:2006.13243}

\bibitem[\protect\citeauthoryear{{Hirano} et~al.,}{{Hirano}
  et~al.}{2020b}]{2020ApJ...890L..27H}
{Hirano} T.,  et~al., 2020b, \mn@doi [\apjl] {10.3847/2041-8213/ab74dc}, \href
  {https://ui.adsabs.harvard.edu/abs/2020ApJ...890L..27H} {890, L27}

\bibitem[\protect\citeauthoryear{{Kastner}, {Principe}, {Punzi}, {Stelzer},
  {Gorti}, {Pascucci}  \& {Argiroffi}}{{Kastner} et~al.}{2016}]{Kastner2016}
{Kastner} J.~H.,  {Principe} D.~A.,  {Punzi} K.,  {Stelzer} B.,  {Gorti} U.,
  {Pascucci} I.,   {Argiroffi} C.,  2016, \mn@doi [\aj]
  {10.3847/0004-6256/152/1/3}, \href
  {http://adsabs.harvard.edu/abs/2016AJ....152....3K} {152, 3}

\bibitem[\protect\citeauthoryear{{Kockarts}}{{Kockarts}}{2002}]{Kockarts2002}
{Kockarts} G.,  2002, \mn@doi [Annales Geophysicae]
  {10.5194/angeo-20-585-2002}, \href
  {https://ui.adsabs.harvard.edu/abs/2002AnGeo..20..585K} {20, 585}

\bibitem[\protect\citeauthoryear{{Kotani} et~al.,}{{Kotani}
  et~al.}{2018}]{Kotani2018}
{Kotani} T.,  et~al., 2018, in \procspie. p. 1070211,
  \mn@doi{10.1117/12.2311836}

\bibitem[\protect\citeauthoryear{{Kubyshkina} et~al.,}{{Kubyshkina}
  et~al.}{2018}]{Kubyshkina2018}
{Kubyshkina} D.,  et~al., 2018, \mn@doi [\apjl] {10.3847/2041-8213/aae586},
  \href {https://ui.adsabs.harvard.edu/abs/2018ApJ...866L..18K} {866, L18}

\bibitem[\protect\citeauthoryear{{Li} \& {Winn}}{{Li} \&
  {Winn}}{2016}]{LiG2016}
{Li} G.,  {Winn} J.~N.,  2016, \mn@doi [\apj] {10.3847/0004-637X/818/1/5},
  \href {https://ui.adsabs.harvard.edu/abs/2016ApJ...818....5L} {818, 5}

\bibitem[\protect\citeauthoryear{{Linsky}, {France}  \& {Ayres}}{{Linsky}
  et~al.}{2013}]{Linsky2013}
{Linsky} J.~L.,  {France} K.,   {Ayres} T.,  2013, \mn@doi [\apj]
  {10.1088/0004-637X/766/2/69}, \href
  {https://ui.adsabs.harvard.edu/abs/2013ApJ...766...69L} {766, 69}

\bibitem[\protect\citeauthoryear{{Linsky}, {Fontenla}  \& {France}}{{Linsky}
  et~al.}{2014}]{Linsky2014}
{Linsky} J.~L.,  {Fontenla} J.,   {France} K.,  2014, \mn@doi [\apj]
  {10.1088/0004-637X/780/1/61}, \href
  {https://ui.adsabs.harvard.edu/abs/2014ApJ...780...61L} {780, 61}

\bibitem[\protect\citeauthoryear{{Luger} \& {Barnes}}{{Luger} \&
  {Barnes}}{2015}]{Luger2015}
{Luger} R.,  {Barnes} R.,  2015, \mn@doi [Astrobiology]
  {10.1089/ast.2014.1231}, \href
  {https://ui.adsabs.harvard.edu/abs/2015AsBio..15..119L} {15, 119}

\bibitem[\protect\citeauthoryear{{Mann} et~al.,}{{Mann}
  et~al.}{2016}]{Mann2016a}
{Mann} A.~W.,  et~al., 2016, \mn@doi [\apj] {10.3847/0004-637X/818/1/46}, \href
  {http://adsabs.harvard.edu/abs/2016ApJ...818...46M} {818, 46}

\bibitem[\protect\citeauthoryear{{Mann} et~al.,}{{Mann}
  et~al.}{2018}]{Mann2018}
{Mann} A.~W.,  et~al., 2018, \mn@doi [\aj] {10.3847/1538-3881/aa9791}, \href
  {http://adsabs.harvard.edu/abs/2018AJ....155....4M} {155, 4}

\bibitem[\protect\citeauthoryear{{Mart{\'\i}n}, {Lodieu}, {Pavlenko}  \&
  {B{\'e}jar}}{{Mart{\'\i}n} et~al.}{2018}]{Martin2018}
{Mart{\'\i}n} E.~L.,  {Lodieu} N.,  {Pavlenko} Y.,   {B{\'e}jar} V. J.~S.,
  2018, \mn@doi [\apj] {10.3847/1538-4357/aaaeb8}, \href
  {https://ui.adsabs.harvard.edu/abs/2018ApJ...856...40M} {856, 40}

\bibitem[\protect\citeauthoryear{{Matsumura}, {Takeda}  \& {Rasio}}{{Matsumura}
  et~al.}{2008}]{Matsumura2008}
{Matsumura} S.,  {Takeda} G.,   {Rasio} F.~A.,  2008, \mn@doi [\apjl]
  {10.1086/592818}, \href
  {https://ui.adsabs.harvard.edu/abs/2008ApJ...686L..29M} {686, L29}

\bibitem[\protect\citeauthoryear{{Matt}, {Brun}, {Baraffe}, {Bouvier}  \&
  {Chabrier}}{{Matt} et~al.}{2015}]{Matt2015}
{Matt} S.~P.,  {Brun} A.~S.,  {Baraffe} I.,  {Bouvier} J.,   {Chabrier} G.,
  2015, \mn@doi [\apjl] {10.1088/2041-8205/799/2/L23}, \href
  {https://ui.adsabs.harvard.edu/abs/2015ApJ...799L..23M} {799, L23}

\bibitem[\protect\citeauthoryear{{McCully} et~al.,}{{McCully}
  et~al.}{2018}]{McCully2018}
{McCully} C.,  et~al., 2018, {Lcogt/Banzai: Initial Release},
  \mn@doi{10.5281/zenodo.1257560}

\bibitem[\protect\citeauthoryear{{Muirhead} et~al.,}{{Muirhead}
  et~al.}{2015}]{Muirhead2015}
{Muirhead} P.~S.,  et~al., 2015, \mn@doi [\apj] {10.1088/0004-637X/801/1/18},
  \href {http://adsabs.harvard.edu/abs/2015ApJ...801...18M} {801, 18}

\bibitem[\protect\citeauthoryear{{Mulders}, {Pascucci}  \& {Apai}}{{Mulders}
  et~al.}{2015}]{Mulders2015}
{Mulders} G.~D.,  {Pascucci} I.,   {Apai} D.,  2015, \mn@doi [\apj]
  {10.1088/0004-637X/798/2/112}, \href
  {http://adsabs.harvard.edu/abs/2015ApJ...798..112M} {798, 112}

\bibitem[\protect\citeauthoryear{{Oklop{\v c}i{\'c}} \& {Hirata}}{{Oklop{\v
  c}i{\'c}} \& {Hirata}}{2018}]{Oklopvcic2018}
{Oklop{\v c}i{\'c}} A.,  {Hirata} C.~M.,  2018, \mn@doi [\apjl]
  {10.3847/2041-8213/aaada9}, \href
  {http://adsabs.harvard.edu/abs/2018ApJ...855L..11O} {855, L11}

\bibitem[\protect\citeauthoryear{{Oklop{\v{c}}i{\'c}}}{{Oklop{\v{c}}i{\'c}}}{2019}]{Oklopvcic2019}
{Oklop{\v{c}}i{\'c}} A.,  2019, arXiv e-prints, \href
  {https://ui.adsabs.harvard.edu/abs/2019arXiv190302576O} {p. arXiv:1903.02576}

\bibitem[\protect\citeauthoryear{{Owen}}{{Owen}}{2019}]{Owen2019}
{Owen} J.~E.,  2019, \mn@doi [Annual Review of Earth and Planetary Sciences]
  {10.1146/annurev-earth-053018-060246}, \href
  {https://ui.adsabs.harvard.edu/abs/2019AREPS..47...67O} {47, 67}

\bibitem[\protect\citeauthoryear{{Parviainen}}{{Parviainen}}{2015}]{Parvianen2015}
{Parviainen} H.,  2015, \mn@doi [\mnras] {10.1093/mnras/stv894}, \href
  {https://ui.adsabs.harvard.edu/abs/2015MNRAS.450.3233P} {450, 3233}

\bibitem[\protect\citeauthoryear{{Riley}}{{Riley}}{2017}]{Riley2017}
{Riley} A.,  2017, {STIS Instrument Hanbook for Cycle 25, Version 16.0}

\bibitem[\protect\citeauthoryear{{Shematovich}, {Ionov}  \&
  {Lammer}}{{Shematovich} et~al.}{2014}]{Shematovich2014}
{Shematovich} V.~I.,  {Ionov} D.~E.,   {Lammer} H.,  2014, \mn@doi [\aap]
  {10.1051/0004-6361/201423573}, \href
  {https://ui.adsabs.harvard.edu/abs/2014A&A...571A..94S} {571, A94}

\bibitem[\protect\citeauthoryear{{Tamura} et~al.,}{{Tamura}
  et~al.}{2012}]{2012SPIE.8446E..1TT}
{Tamura} M.,  et~al., 2012, in \procspie. p. 84461T, \mn@doi{10.1117/12.925885}

\bibitem[\protect\citeauthoryear{{Thao} et~al.,}{{Thao}
  et~al.}{2020}]{Thao2020}
{Thao} P.~C.,  et~al., 2020, \mn@doi [\aj] {10.3847/1538-3881/ab579b}, \href
  {https://ui.adsabs.harvard.edu/abs/2020AJ....159...32T} {159, 32}

\bibitem[\protect\citeauthoryear{{Tian}, {France}, {Linsky}, {Mauas}  \&
  {Vieytes}}{{Tian} et~al.}{2014}]{Tian2014}
{Tian} F.,  {France} K.,  {Linsky} J.~L.,  {Mauas} P. J.~D.,   {Vieytes} M.~C.,
   2014, \mn@doi [Earth and Planetary Science Letters]
  {10.1016/j.epsl.2013.10.024}, \href
  {https://ui.adsabs.harvard.edu/abs/2014E&PSL.385...22T} {385, 22}

\bibitem[\protect\citeauthoryear{{Tody}}{{Tody}}{1986}]{Tody1986}
{Tody} D.,  1986, in {Crawford} D.~L.,  ed.,  \procspie Vol. 627,
  Instrumentation in astronomy VI. p.~733, \mn@doi{10.1117/12.968154}

\bibitem[\protect\citeauthoryear{{Triaud}}{{Triaud}}{2017}]{Triaud2017}
{Triaud} A.~H.~M.~J.,  2017, {The Rossiter-McLaughlin Effect in Exoplanet
  Research}.
Springer, p.~2, \mn@doi{10.1007/978-3-319-30648-3_2-1}

\bibitem[\protect\citeauthoryear{{Virtanen} et~al.,}{{Virtanen}
  et~al.}{2019}]{Scipy2019}
{Virtanen} P.,  et~al., 2019, arXiv e-prints, \href
  {https://ui.adsabs.harvard.edu/abs/2019arXiv190710121V} {p. arXiv:1907.10121}

\bibitem[\protect\citeauthoryear{{Watson}, {Donahue}  \& {Walker}}{{Watson}
  et~al.}{1981}]{Watson1981}
{Watson} A.~J.,  {Donahue} T.~M.,   {Walker} J.~C.~G.,  1981, \mn@doi [\icarus]
  {10.1016/0019-1035(81)90101-9}, \href
  {https://ui.adsabs.harvard.edu/abs/1981Icar...48..150W} {48, 150}

\bibitem[\protect\citeauthoryear{{Wright}, {Newton}, {Williams}, {Drake}  \&
  {Yadav}}{{Wright} et~al.}{2018}]{Wright2018}
{Wright} N.~J.,  {Newton} E.~R.,  {Williams} P. K.~G.,  {Drake} J.~J.,
  {Yadav} R.~K.,  2018, \mn@doi [\mnras] {10.1093/mnras/sty1670}, \href
  {https://ui.adsabs.harvard.edu/abs/2018MNRAS.479.2351W} {479, 2351}

\bibitem[\protect\citeauthoryear{{Yasui} et~al.,}{{Yasui}
  et~al.}{2019}]{Yasui2019}
{Yasui} C.,  et~al., 2019, \mn@doi [\apj] {10.3847/1538-4357/ab45ee}, \href
  {https://ui.adsabs.harvard.edu/abs/2019ApJ...886..115Y} {886, 115}

\bibitem[\protect\citeauthoryear{{Youngblood} et~al.,}{{Youngblood}
  et~al.}{2016}]{Youngblood2016}
{Youngblood} A.,  et~al., 2016, \mn@doi [\apj] {10.3847/0004-637X/824/2/101},
  \href {http://adsabs.harvard.edu/abs/2016ApJ...824..101Y} {824, 101}

\makeatother
\end{thebibliography}


\bsp	
\label{lastpage}
\end{document}